\documentclass[12pt,epsf]{article}
\usepackage{graphicx}

\newcommand{\be}{\begin{equation}}
\newcommand{\ee}{\end{equation}}
\newcommand{\bea}{\begin{eqnarray}}
\newcommand{\eea}{\end{eqnarray}}
\newcommand{\beas}{\begin{eqnarray*}}
\newcommand{\eeas}{\end{eqnarray*}}
\newcommand{\ba}{\begin{array}}
\newcommand{\ea}{\end{array}}

\newcommand{\tr}{\mathrm{Tr}}

\newcommand{\nbox}{{\,\lower0.9pt\vbox{\hrule \hbox{\vrule height 0.2 cm \hskip 0.19 cm \vrule height 0.2 cm}\hrule}\,}}

\def\href#1#2{#2}

\begin{document}
\begin{titlepage}
\hfill

\vspace*{20mm}
\begin{center}
{\Large \bf Building up spacetime with quantum entanglement}

\vspace*{15mm}
\vspace*{1mm}

Mark Van Raamsdonk

\vspace*{1cm}

{Department of Physics and Astronomy,
University of British Columbia\\
6224 Agricultural Road,
Vancouver, B.C., V6T 1W9, Canada\\
{\it mav@phas.ubc.ca}
}

\vspace*{1cm}
\end{center}

\begin{abstract}
In this essay based on 0907.2939 \cite{VanRaamsdonk:2009ar}, we argue that the emergence of classically connected spacetimes is intimately related to the quantum entanglement of degrees of freedom in a non-perturbative description of quantum gravity. Disentangling the degrees of freedom associated with two regions of spacetime results in these regions pulling apart and pinching off from each other in a way that can be quantified by standard measures of entanglement.

\end{abstract}
\vskip 2cm

\begin{center}
Essay written for the Gravity Research Foundation \\ 2010 Awards for Essays on Gravitation
\vskip 1cm

March 31, 2010
\end{center}
\end{titlepage}

\vskip 1cm
\vskip 0.1 in
\noindent
{\bf Introduction}
\vskip 0.1 in
\noindent
The gravity / gauge theory correspondence \cite{bfss, malda,agmoo} in string theory represents exciting progress towards finding a general non-perturbative description of quantum gravity. It posits that certain quantum gravitational theories with fixed spacetime asymptotic behavior are exactly equivalent to ordinary quantum field theories. We can view this correspondence as providing a complete non-perturbative definition of the quantum gravity theory via a quantum field theory. However, despite a great deal of evidence for the validity of this correspondence, we do not have a deep understanding of why or how spacetime/gravity emerges from the degrees of freedom of the field theory.

In this essay, we will argue, based on widely accepted examples of gauge theory / gravity duality, that the emergence of spacetime in the gravity picture is intimately related to the quantum entanglement of degrees of freedom in the corresponding conventional quantum system. We will begin by showing that certain quantum superpositions of states corresponding to disconnected spacetimes give rise to states that are interpreted as classically connected spacetimes.  More quantitatively, we will see in a simple example that decreasing the entanglement between two sets of degrees of freedom (e.g. by continuously varying the quantum state in the field theory description) effectively increases the proper distance between the corresponding spacetime regions, while decreasing the area separating the two regions.
\vskip 0.1 in
\noindent
{\bf Classical connectivity from quantum superposition}
\vskip 0.1 in
\noindent
The most familiar example of gauge-theory / gravity duality involves an equivalence between conformal field theories (CFTs) and asymptotically anti-de Sitter (AdS) spacetimes. For specific CFTs, each state of the field theory on a sphere ($\times$ time) corresponds to a spacetime in some specific theory of quantum gravity where the asymptotics of the spacetime are the same as global AdS spacetime.

Let us now consider a slightly more complicated example, where we build a larger quantum system by taking two (non-interacting) copies of our conformal field theory on $S^d$. As usual, the Hilbert space for this system will be the tensor product ${\cal H} = {\cal H}_1 \otimes {\cal H}_2$ of the Hilbert spaces for the component systems.

In our new system, the simplest quantum states to consider are product states (i.e. states with no entanglement between the two subsystems)
\[
|\Psi \rangle = |\Psi_1 \rangle \otimes |\Psi_2 \rangle \; .
\]
It is easy to provide a gravity interpretation for such a state. Since the degrees of freedom of the two CFTs do not interact in any way, and since there is no entanglement between the degrees of freedom for this state, the interpretation must be that we have two completely separate physical systems. If $|\Psi_1 \rangle$ is dual to one asymptotically AdS spacetime and $|\Psi_2 \rangle$ is dual to some other spacetime, the product state is dual to the disconnected pair of spacetimes.

We next consider a state in which the two subsystems are entangled. If $| E_i \rangle$ represents the $i$th energy eigenstate for a single CFT on $S^d$, let us define the state
\be
\label{hhstate}
|\psi (\beta) \rangle = \sum_i e^{- \beta E_i \over 2} | E_i \rangle \otimes | E_i \rangle
\ee
This state is a sum of product states $| E_i \rangle \otimes | E_i \rangle$. Since we just argued that each of these product states should be interpreted on the gravity side as a spacetime with two disconnected components, the literal interpretation of the state $|\psi (\beta) \rangle$ is that it is a quantum superposition of disconnected spacetimes.
\begin{figure}
\centering
\includegraphics[width=\textwidth]{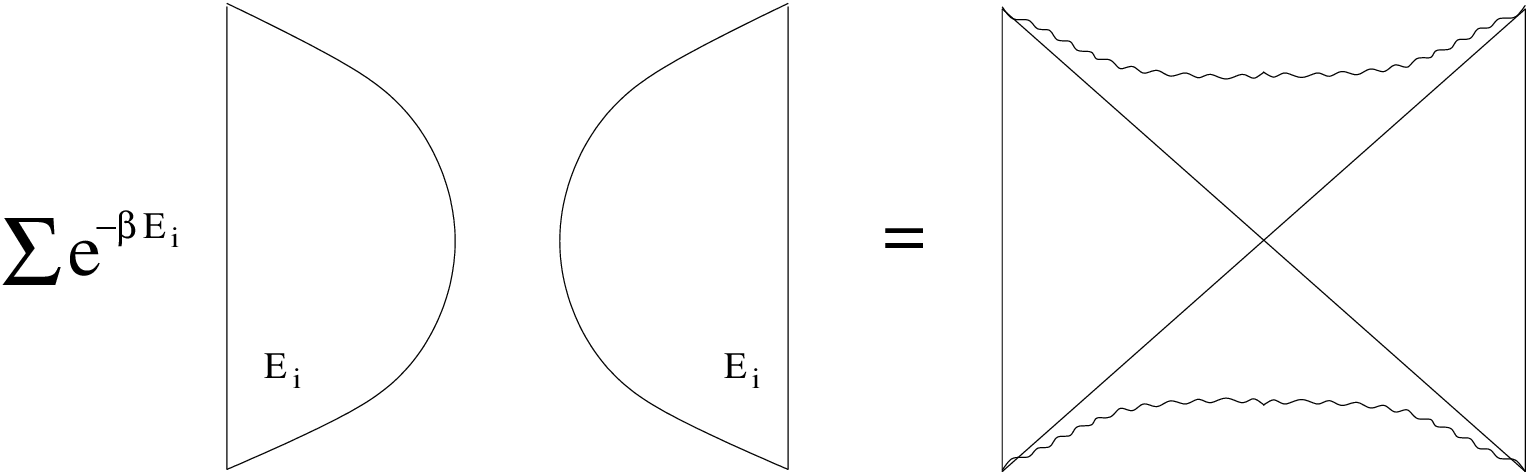}
\caption{Gravity interpretations for the entangled state $|\psi (\beta) \rangle$ in a quantum system defined by a pair of noninteracting CFTs on $S^d$ times time. The diagram on the right is the Penrose diagram for the maximally extended AdS-Schwarzschild black hole. }
\end{figure}
However, it has been argued \cite{eads,hm,bklt} that precisely this state $|\psi (\beta) \rangle$ corresponds to the (connected) eternal AdS black hole spacetime, whose Penrose diagram is sketched in figure 1.

The motivation for this statement is as follows. This is a spacetime with two equivalent asymptotically AdS regions, suggesting that the dual description should involve two copies of the CFT. An observer in either asymptotic region sees the Schwarzschild AdS black hole spacetime, which is understood \cite{wittenthermal} to correspond to the thermal state of a conformal field theory. On the other hand, starting from the state (\ref{hhstate}), and tracing over the degrees of freedom of one of the CFTs, we find that the density matrix for the remaining CFT is exactly the thermal density matrix:
\[
\tr_2(|\psi \rangle \langle \psi |) = \sum_i e^{- \beta E_i } | E_i \rangle \langle E_i | = \rho_T \; .
\]
Furthermore, the presence of horizons in the black hole spacetime which forbid communication between the two asymptotic regions may be naturally associated with the absence of interactions between the two CFTs. Thus, the state $|\psi (\beta) \rangle$ has properties which are completely consistent with its interpretation as the eternal AdS black hole.

If this identification is correct, we have a remarkable conclusion: the state $|\psi (\beta) \rangle$ which clearly represents a quantum superposition of disconnected spacetimes may also be identified with a classically connected spacetime. In this example, {\it classical connectivity arises by entangling the degrees of freedom in the two components}. In the next section, we will try to test the idea that emergent spacetimes in gauge-theory / gravity duality are built up by entangling degrees of freedom in the non-perturbative description.
\vskip 0.1 in
\noindent
{\bf A disentangling experiment}
\vskip 0.1 in
\noindent
Let us return to the simpler case of a single CFT on $S^d$. We would like to do a thought experiment in which we start with the vacuum state of the field theory, dual to gravity on pure global AdS spacetime, and see what happens to the dual geometry when we gradually change the state to disentangle some of the degrees of freedom. To be specific, we divide the sphere into two parts (e.g. hemispheres) which we label $A$ and $B$.

Since the CFT is a local quantum field theory, there are specific degrees of freedom associated with specific spatial regions, so we can decompose the Hilbert space ${\cal H} = {\cal H}_A \otimes {\cal H}_B$. A simple quantitative measure of the entanglement between $A$ and $B$ is the entanglement entropy \cite{nc}, defined to be the von Neumann entropy
\[
S(A) = - \tr( \rho_A \log \rho_A)
\]
of the density matrix for the subsystem $A$,
\[
\rho_A = \tr_{B}(|\Psi \rangle \langle \Psi |) \; .
\]
This is typically a divergent quantity, but we can consider a field theory defined with a cutoff (e.g. on a lattice), such that the entanglement entropy is finite.
\begin{figure}
\centering
\includegraphics[width=0.25\textwidth]{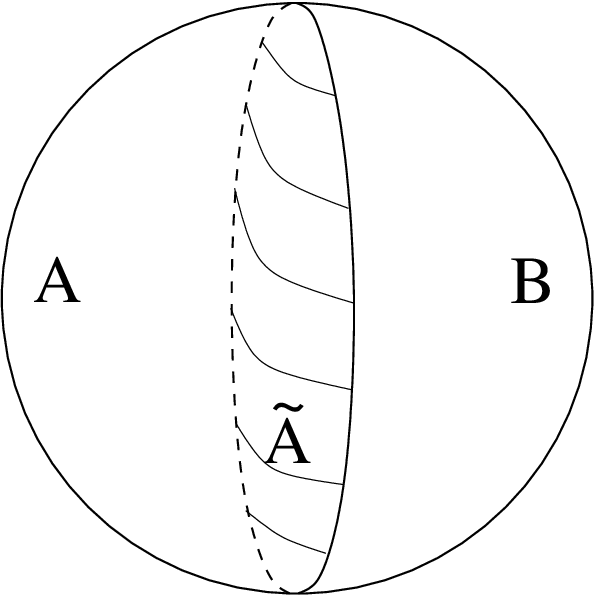}
\caption{According to \cite{rt}, the entanglement entropy $S(A)$ between regions A and B in the field theory is related to the area of the minimal surface $\tilde{A}$ in the dual geometry such that the boundary of $\tilde{A}$ coincides with the boundary of $A$: $S(A) = {\rm Area}(\tilde{A})/(4 G_N)$. In the diagram, spatial geometry in the gravity picture is represented by the interior of the ball, while the geometry on which the field theory lives is identified with the boundary sphere.}
\end{figure}
Now, starting with the vacuum state, we can ask what happens to the dual spacetime when we vary the quantum state in such a way that the entanglement entropy $S(A)$ decreases. Using the recent proposal of Ryu and Takayanagi \cite{rt}, we can make a very precise statement about what happens: the area of the minimal surface $\tilde{A}$ in the dual spacetime which separates the spherical boundary into its two components $A$ and $B$ decreases, in direct proportionality to the decrease in entanglement entropy (see figure 2). Since the surface $\tilde{A}$ is a dividing surface between two regions of the dual space, we see that as entanglement is decreased to zero, the two regions of space are pinching off from each other.\footnote{Here and below, we should keep in mind that the spacetime will likely cease to have a completely geometrical description before the entanglement is strictly zero.}

Another insight into the behavior of the dual geometry as we decrease entanglement comes from considering the {\it mutual information} \cite{nc} between any two subsystems $C \subset A$ and $D \subset B$, defined to be
\[
I(C,D) = S(C) + S(D) - S(C \cup D) \; .
\]
It can be shown that $I(C,D)$ is always non-negative, and zero if and only if the density matrix for $C \cup D$ is the tensor product of the density matrices for $C$ and $D$. As the entanglement between $A$ and $B$ decreases to zero, the mutual information between any two regions $C$ and $D$ goes to zero also.

\begin{figure}
\centering
\includegraphics[width=0.25\textwidth]{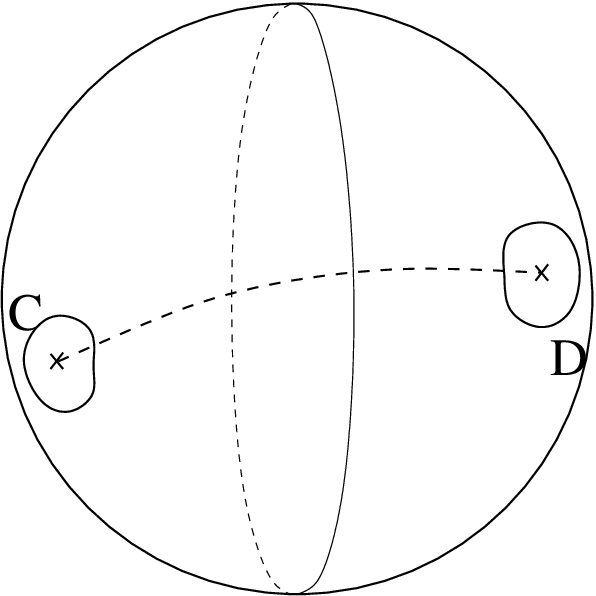}
\caption{Length $L$ of geodesic (dashed line) connecting boundary points in $C$ and $D$ must go to infinity if mutual information between C and D decreases to zero.}
\end{figure}

From the geometrical point of view, we will now argue that this decrease in mutual information between $C$ and $D$ implies an increase the proper distance between the corresponding regions near the boundary of spacetime. First, we note that mutual information provides an upper bound on correlations in a system. It is straightforward to prove \cite{wvhc} that for any operators ${\cal O}_C$ and ${\cal O}_D$, acting only on the subsystems $C$ and $D$, we have
\be
\label{corrbound}
I(C,D) \ge \frac{(\langle {\cal O}_C {\cal O}_D \rangle - \langle {\cal O}_C \rangle \langle {\cal O}_D \rangle  )^2}{2|{\cal O}_C|^2 |{\cal O}_D|^2} \; .
\ee
Thus, if we continuously vary a state such that the mutual information between $C$ and $D$ go to zero, then all correlations must decrease to zero also. In the context of AdS/CFT, certain two-point correlators of local operators (those dual to very massive particles in the dual spacetime) provide a direct measure of the proper distance through the spacetime between the boundary points at which the operators are inserted. Specifically, we have:
\be
\label{corr}
\langle {\cal O}_C(x_C) {\cal O}_D (x_D) \rangle \sim e^{-m L}
\ee
where $m$ is the mass and $L$ is the length of the shortest geodesic connecting $x_C$ and $x_D$ (again, we can work in a field theory with explicit cutoff so everything is finite). Combining (\ref{corr}) and (\ref{corrbound}), we see that as the entanglement between degrees of freedom in region $A$ and region $B$ (and therefore the mutual information $I(C,D)$) drops to zero, the length of the shortest bulk path between the points $x_C$ and $x_D$ must go to infinity (figure 3).
\begin{figure}
\centering
\includegraphics[width=0.3\textwidth]{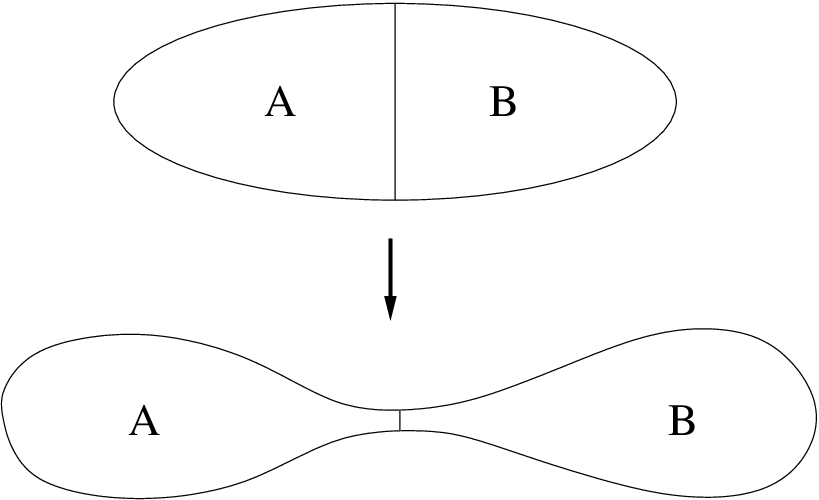}
\caption{Effect on geometry of decreasing entanglement between holographic degrees of freedom corresponding to $A$ and $B$: area separating corresponding spatial regions decreases while distance between points increases. The boundary geometry remains fixed (despite appearances in the diagram).}
\end{figure}
Together with the result of the previous subsection, we obtain the following picture. As the entanglement between two sets of degrees of freedom in a nonperturbative description of quantum gravity drops to zero, the proper distance between the corresponding spacetime regions goes to infinity, while the area of the minimal surface separating the regions decreases to zero. Roughly speaking, the two regions of spacetime pull apart and pinch off from each other, as shown in figure 4.
\begin{figure}
\centering
\includegraphics[width=0.3\textwidth]{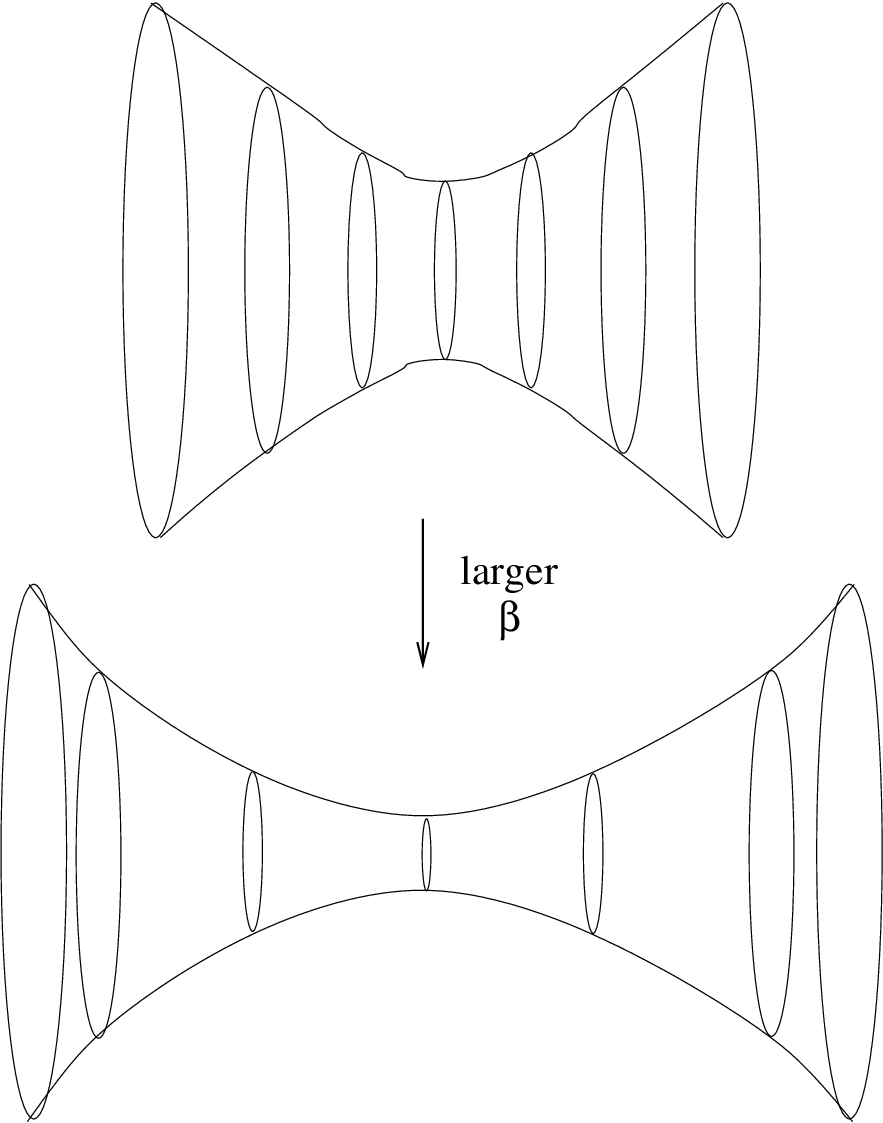}
\caption{Spatial section of eternal black hole for two different temperatures (corresponding to a horizontal line through the middle of the Penrose diagram of figure 1). For low temperature (large $\beta$), where entanglement between the two CFTs is smaller, the asymptotic regions are further apart and separated by a surface of smaller area.}
\end{figure}
As seen in figure 5, these quantitative features can be seen explicitly in the example of the eternal AdS black hole, where we can decrease the entanglement between the two CFTs by increasing the inverse temperature parameter $\beta$.
\vskip 0.1 in
\noindent
{\bf Conclusions}
\vskip 0.1 in
\noindent
We have seen that we can connect up spacetimes by entangling degrees of freedom and tear them apart by disentangling. It is fascinating that the intrinsically quantum phenomenon of entanglement appears to be crucial for the emergence of classical spacetime geometry.

\section*{Acknowledgements}
This work has been supported in part by the Natural Sciences and Engineering Research Council of Canada, the Alfred P. Sloan Foundation, and the Canada Research Chairs programme.

\end{document}